\documentclass{PoS}

\usepackage{soul,xcolor}
\setstcolor{red}

\definecolor{hookgreen}{rgb}{0.0,0.44,0.0}

\title{A new calculation of Earth-skimming very- and ultra-high energy tau neutrinos }

\ShortTitle{Earth-skimming tau neutrinos}

\author{\speaker{Mary Hall Reno}\\
        Department of Physics and Astronomy, University of Iowa, Iowa City, IA 52242, USA\\
        E-mail: \email{mary-hall-reno@uiowa.edu}}

\author{Tonia M. Venters,$^1$ John F. Krizmanic,$^{2,3}$ Luis A. Anchordoqui,$^{4,5,6}$ Claire Gu\'epin,$^7$ Angela V. Olinto$^8$\\
        $^1$Astrophysics Science Division, NASA Goddard Space Flight Center, Greenbelt, MD 20771, USA; $^2$CRESST/ NASA Goddard Space Flight Center, Greenbelt, MD 20771, USA; $^3$University of  Maryland, Baltimore County,
        Baltimore, MD 21250, USA;
        $^4$Department of Physics, Graduate Center, City University of New York, NY 10016, USA; $^5$Department of Physics and Astronomy, Lehman College (CUNY), NY 10468, USA; $^6$Department of Astrophysics, American Museum of Natural History, NY 10024, USA; 
         $^7$Sorbonne Universit\'e, CRNS, UMR 7095,
        Institute d'Astrophysique de Paris, 98 bis bd Arago, 75014 Paris, France;
        $^8$Department of Astronomy \& Astrophysics, KICP, EFI, The University of Chicago, Chicago, IL 60637, USA}
        
\author{for the POEMMA Collaboration}

\abstract{Cosmic neutrinos above a PeV are produced either within astrophysical sources or when ultra-high energy cosmic rays interact in transit through the cosmic background radiation. Detection of these neutrinos will be essential for  understanding cosmic ray acceleration, composition and source evolution.  By using the Earth as a tau neutrino converter for upward-going extensive air showers from tau decays, balloon-borne and space-based instruments can take advantage of a large volume and mass of the terrestrial neutrino target. The theoretical inputs and uncertainties in determining the tau lepton exit probabilities and their translation to detection acceptance will be discussed in the context of a new calculation we have performed.  We quantify the experimental detection capability based on our calculation, including using the Probe of Extreme Multi-Messenger Astrophysics (POEMMA) concept study response parameters for optical air Cherenkov detection. These case studies are used to illustrate the features and uncertainties in upward tau air shower detection.}

\FullConference{36th International Cosmic Ray Conference -ICRC2019-\\
		July 24th - August 1st, 2019\\
		Madison, WI, U.S.A.}

\begin{document}

\section{Introduction}

The detection of ultrahigh energy neutrinos from individual astrophysical sources or the diffuse astrophysical neutrino flux from all sources will help us untangle the conditions that enable cosmic ray acceleration in astrophysical environments and the cosmological evolution of their sources \cite{Kotera:2011cp,Anchordoqui:2018qom}. Neutrinos from cosmic ray interactions with the cosmic background radiation will additionally reveal information about cosmic ray composition. Neutrino signals may accompany electromagnetic and gravitational messengers, or in some cases, they may be the primary signal at the site of cosmic ray acceleration.
The detection of cosmic neutrinos above energies of $10^{16}$ eV$=10$ PeV has not yet been achieved, but it is a goal of many instruments with neutrino detection capabilities. Instruments like IceCube 
\cite{Aartsen:2018vtx},  ANTARES and KM3NeT \cite{Sanguineti:2019pkv},
the Askaryan Radio Array (ARA) \cite{Allison:2011wk} and the Antarctic Ross Ice-Shelf Antenna Neutrino Array (ARIANNA) \cite{Barwick:2014pca} take advantage of large volumes of ice or water to act as neutrino targets. IceCube and the Pierre Auger Observatory have set
upper limits on the high energy diffuse neutrino flux from searches of Earth-skimming neutrino interactions as well as  cascades on ice  and deep horizontal air showers \cite{Aartsen:2018vtx,Aab:2019auo,Zas:2017xdj,Abreu:2012zz}. The Antarctic Impulsive Transient Antenna (ANITA) radio antennas on balloon missions over the South Pole has set the most restrictive upper limits on the diffuse neutrino flux at the highest energies \cite{Gorham:2019guw}. 
Thus far, at energies above $5\times 10^6$ GeV, limits but not observations of the diffuse astrophysical neutrino flux have been made. 

Neutrino messengers from transient neutrino sources are a topic of intense study.
In 2018, the IceCube detection of a neutrino event coincident with a gamma-ray flaring blazar  started  a multi-messenger era that includes neutrinos \cite{IceCube:2018dnn}. Larger volumes and accumulated observation time will open up multi-messenger particle astrophysics to neutrino observations.

The Earth-skimming signals of tau neutrinos are the target of the proposed  ground-based detectors Trinity \cite{Otte:2018uxj}  and GRAND \cite{Fang:2017mhl,Alvarez-Muniz:2018bhp}. 
The nearly horizontal or upward-going air showers come from tau decays, where the taus come from tau neutrino charged current interactions in the Earth (see, e.g., Refs. \cite{Fargion:2000iz,Bertou:2001vm,Feng:2001ue}). 
A feature of tau neutrino propagation is that while the initial tau neutrino flux is attenuated by the Earth, as with other neutrino flavors, when the taus they produce decay, tau neutrinos are regenerated,  though at lower energies \cite{Halzen:1998be}.

Satellite-based neutrino detectors will be able to access target volumes for neutrino interactions that are even larger than for existing detectors \cite{Reno:2019jtr}. The Probe of Extreme Multi-Messenger Astrophysics (POEMMA) mission \cite{Olinto:2017xbi,Olinto:2019icrc} is designed for dual detection: of cosmic ray air showers via fluorescence signals and of upward-going air showers from neutrino interactions in the Earth.  The sensitivity of the proposed POEMMA neutrino detectors, based on twin satellites at altitude $h=525$ km flying in formation,  will be to angles from $7^\circ$ below the limb in diffuse neutrino flux mode. POEMMA's coverage of a target-of-opportunity (ToO) sources can extend to $\sim 18^\circ$ below the limb \cite{Venters:2019}. POEMMA will be able to respond to alerts for flaring sources, with a $90^\circ$ slew time of $500$~s \cite{Guepin:2018yuf}. We discuss  below POEMMA's projected sensitivity to tau neutrino fluxes and fluences. More details can be found in Refs. \cite{Reno:2019jtr} and \cite{Venters:2019}.

\section{Effective aperture and effective area}

POEMMA's effective aperture for diffuse neutrino fluxes and its effective area for the neutrino fluence from transient sources depend on tau neutrino interactions that produce taus via the standard model charged current interaction, on tau propagation through materials and tau electromagnetic energy loss, on tau decay probabilities as a function of altitude and the corresponding probability to detect the signal from the extensive air shower (EAS) from that tau decay.  At energies above $10^7$ GeV, the neutrino and antineutrino cross sections are equal, so henceforth, we refer to them both as ``neutrinos.''

\begin{figure}
    \centering
    \includegraphics[width=0.45\columnwidth, trim=0cm -0.5cm 0cm 0cm, clip]{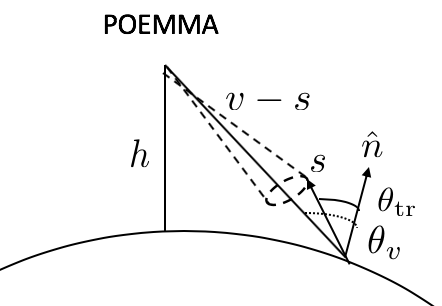}
    \includegraphics[width=0.45\columnwidth]{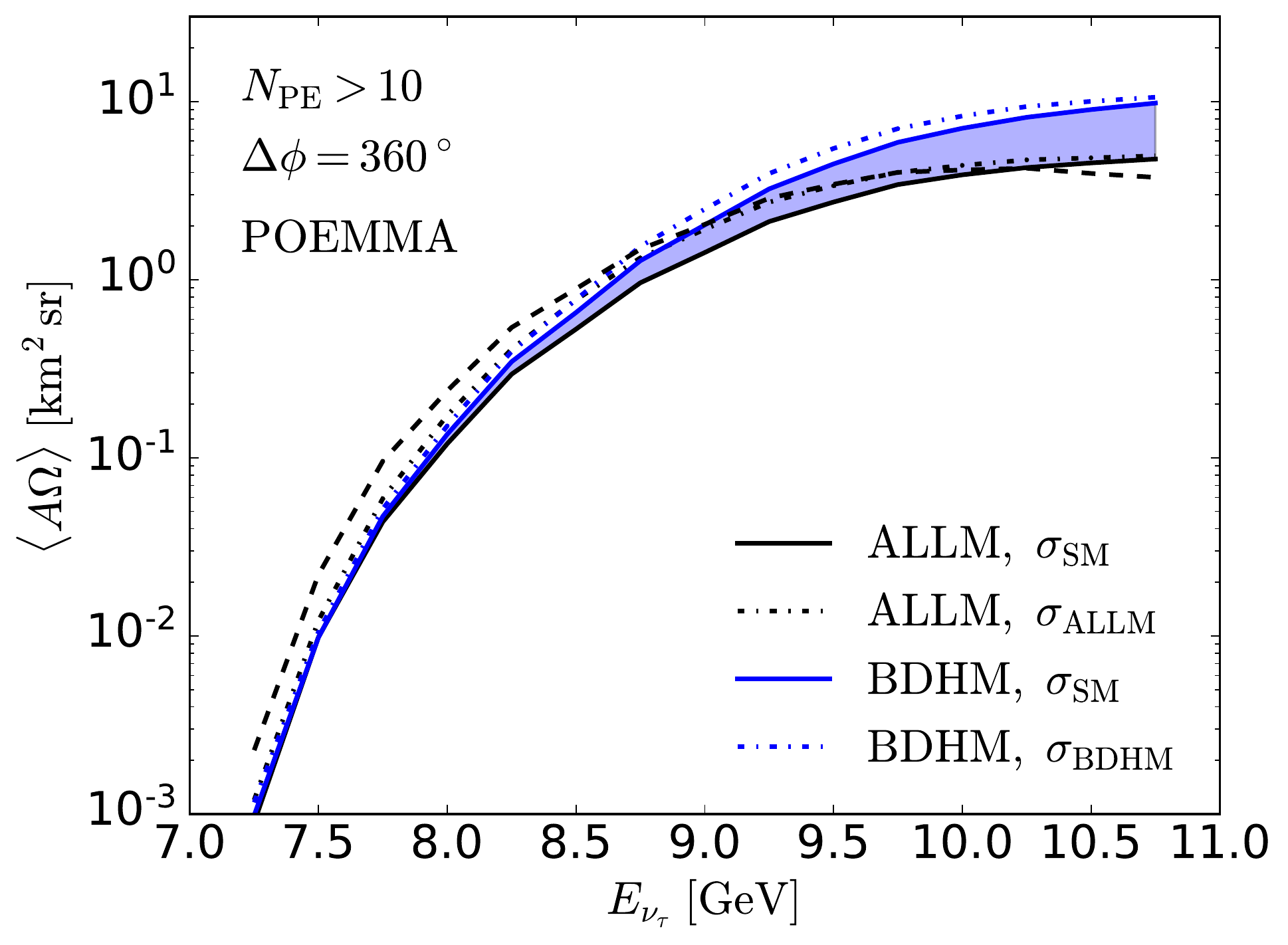}
    \caption{{\it Left}: Definitions of angles and distances for POEMMA, at an altitude $h$ above the surface of the Earth. {\it Right}: The effective aperture for POEMMA at an altitude $h=525$ km, viewing $7^\circ$ below the limb of the Earth. See text for inputs to the solid and dot-dashed lines. The dashed curve comes from using the ALLM tau energy loss model with $\sigma_{\rm SM}$, with rock in the final 3 km layer of the Earth.}
    \label{fig:geometry}
\end{figure}

Figure \ref{fig:geometry} (left) shows our definition of angles 
and distances relevant for a detector an altitude $h$ above the surface of the Earth. The angle $\theta_v$ is the angle relative to the local normal $\hat{n}$ of the line of sight from tau emergence point to the detector, while $\theta_{\rm tr}$ is the angle the exiting tau trajectory makes with respect to the same local normal. The detector is a distance $v$ from the exit point. The tau decays a distance $s$ from the exit point. 

The effective aperture for tau neutrino energy $E_{\nu_\tau}$, relevant for the diffuse flux sensitivity calculation, can be written as \cite{Motloch:2013kva}:
\begin{equation}
\langle A\Omega\rangle  \left(E_{\nu_\tau}\right)= \int_{S}\int_{\Delta\Omega_{\rm tr}} dP_{\rm obs}\ \hat{r}\cdot\hat{n}\, dS \, d\Omega_{\rm tr}\ .
\label{eq:aperture}
\end{equation}
The infinitesimal area element $dS$ is the patch of surface area viewable by the detector. For POEMMA, this is in an annular zone from the limb to $7^\circ$ below the limb. We evaluate the
effective aperture for viewing over the full $\Delta\phi=360^\circ$ azimuthal angle. POEMMA's current design has a coverage of $\Delta\phi = 30^\circ$. The observation probability can be written in terms of exit probability $P_{\rm exit}$, detection
probability $P_{\rm det}$ and the decay probability $p_{\rm decay} ds$ for an infinitesimal path length 
$ds$:
\begin{equation}
dP_{\rm obs} =  ds'\,  P_{\rm exit}(E_{\nu_\tau},\theta_{\rm tr}) 
  {p}_{\rm decay}(s') P_{\rm det}(E_{\nu_\tau},\theta_v,\theta_{\rm tr},s')
\label{eq:pobs}
\end{equation}
Implicit is an integral over tau energies: the exit probability for $\nu_\tau \to \tau$ involves a distribution of energies. Both the decay length and the air shower energy depend on the tau energy. 
The detection probability $P_{\rm det}$ for the upward air shower also depends on whether the detector is within the Cherenkov cone of the EAS from the tau decay. In the diagram in Fig. \ref{fig:geometry} (left) for decay distance $s$, this means the trajectory of the
decay lies within the circle of radius $\sim (v-s)\theta_{\rm Ch}^{\rm eff}$ around the line of sight. The effective Cherenkov angle $\theta_{\rm Ch}^{\rm eff}$ is on the order of $1.5^{\circ}$, though its exact value depends on the altitude of the decay, $\theta_{\rm tr}$, and the shower energy \cite{Reno:2019jtr}.  To first approximation, we can take $\theta_v\simeq \theta_{\rm tr}$. Our detailed evaluation in Ref. \cite{Reno:2019jtr} demonstrates that this is a good approximation.

The photon density at the detector is a function of altitude, $\theta_{\rm tr}$ and the shower energy.  The 2.5 m$^2$ effective optical collection area and quantum efficiency of 0.2 together with the number density of the photons in the Cherenkov cone give the number of photoelectrons. POEMMA's minimum number of photoelectrons for detection is 10 \cite{Reno:2019jtr}. In our evaluations below, we have assumed that the shower energy is half of the tau energy. 
The tau decay probability depends on the tau energy $E_\tau$. For reference, the tau decay length with the $\gamma$-factor $\gamma = E_\tau/(m_\tau c^2)$ is $\gamma c\tau = 5$ km$\times E_\tau/10^8$ GeV.

The exit probability, as noted above, depends on the neutrino charged current cross section and tau electromagnetic energy loss
\cite{Reno:2019jtr,Alvarez-Muniz:2017mpk}. Inputs to both the neutrino cross section and electromagnetic energy loss through photonuclear interactions have uncertainties associated with their high energy extrapolations
\cite{Jeong:2017mzv}.
To illustrate the uncertainties associated with the exit probability, we use two models of tau electromagnetic energy loss: the Abramowicz {\it et al.} (ALLM) \cite{Abramowicz:1991xz,Abramowicz:1997ms} parameterization of the electromagnetic structure function and the Block {\it et al.} (BDHM) \cite{Block:2014kza} extrapolation that predicts less energy loss per unit column depth than the ALLM choice. The solid lines in
Fig. \ref{fig:geometry} (right) shows the effective aperture for these two energy loss models, with a common neutrino cross section based on a parton distribution function evaluation. The shaded blue band between the solid curves gives an estimate of the energy loss uncertainties. At lower energies, the ALLM and BDHM parameterizations are similar, but at higher energies, differences are as much as a factor of 
$\sim 2$ at $E_{\nu_\tau}\simeq 5\times 10^{10}$ GeV. The dot-dashed lines show the effective aperture
when the neutrino cross sections are modified to have the same high energy extrapolations as the photonuclear energy loss parameters, namely with a high energy behavior governed by the ALLM or BDHM extrapolation. There is $<46\%$ difference between the ALLM evaluation with neutrino cross section $\sigma_{\rm SM}$ and $\sigma_{\rm ALLM}$. The difference is smaller, $<23\%$,  comparing neutrino interactions with 
 $\sigma_{\rm SM}$ and $\sigma_{\rm BDHM}$.
 
 The advantage of detection of upward-going EAS over water compared to over rock has been discussed in the literature \cite{PalomaresRuiz:2005xw,Alvarez-Muniz:2017mpk}. Our default model of density layers in the Earth has an outermost layer of 3 km of water. The dashed line in Fig. \ref{fig:geometry} (right) shows the effective aperture if the outermost layer is standard rock with mass density $\rho=2.65$
 g/cm$^3$. Water has a slight advantage at the highest energies, while rock, with more target nucleons for neutrino interactions within $\gamma c\tau$ of the Earth's surface at lower energies, is more advantageous.
 
 For the sensitivity to target-of-opportunity transient neutrino flares, the effective area viewable by POEMMA is required. The effective area is approximately
 \begin{equation}
 \label{eq:aeff}
 A(\theta_{\rm tr},E_{\nu_\tau})\simeq \int dP_{\rm obs} \times \pi (v-s)^2(\theta_{\rm Ch}^{\rm eff})^2\ ,
 \end{equation}
 where $\pi(v-s)^2(\theta_{\rm Ch}^{\rm eff})^2$ is the area of the disk in Fig. \ref{fig:geometry} (left) that is viewable for a given angle $\theta_{\rm tr}\simeq \theta_v$. Our results for point sources rely on Eq. (\ref{eq:aeff}).

 \section{$\nu_\tau\to \tau$ fluxes and the sensitivity of POEMMA}
 
\begin{figure}
    \centering
    \includegraphics[width=0.45\columnwidth]{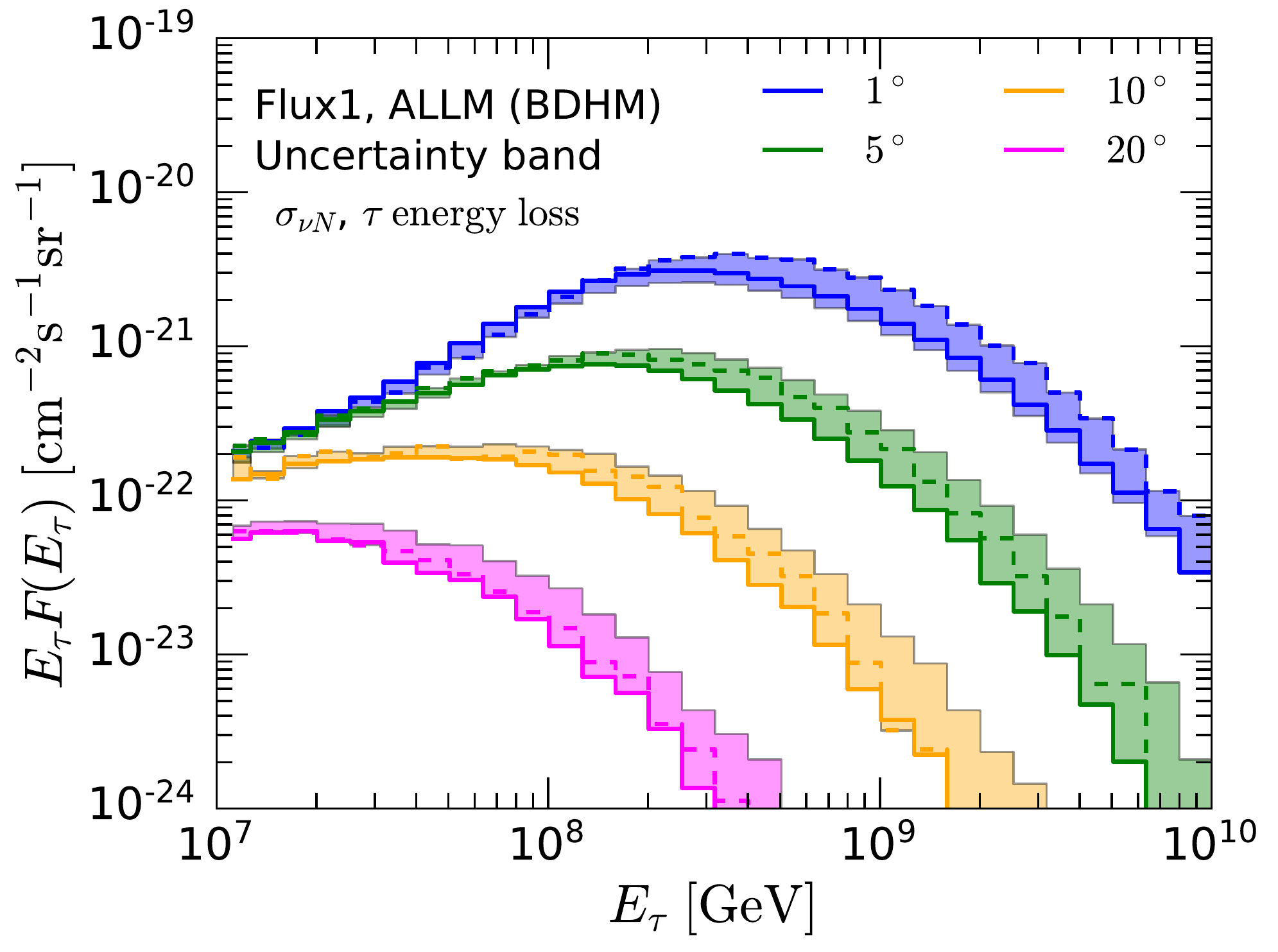}
        \includegraphics[width=0.45\columnwidth, trim=7cm 3.5cm 4cm 3cm, clip]{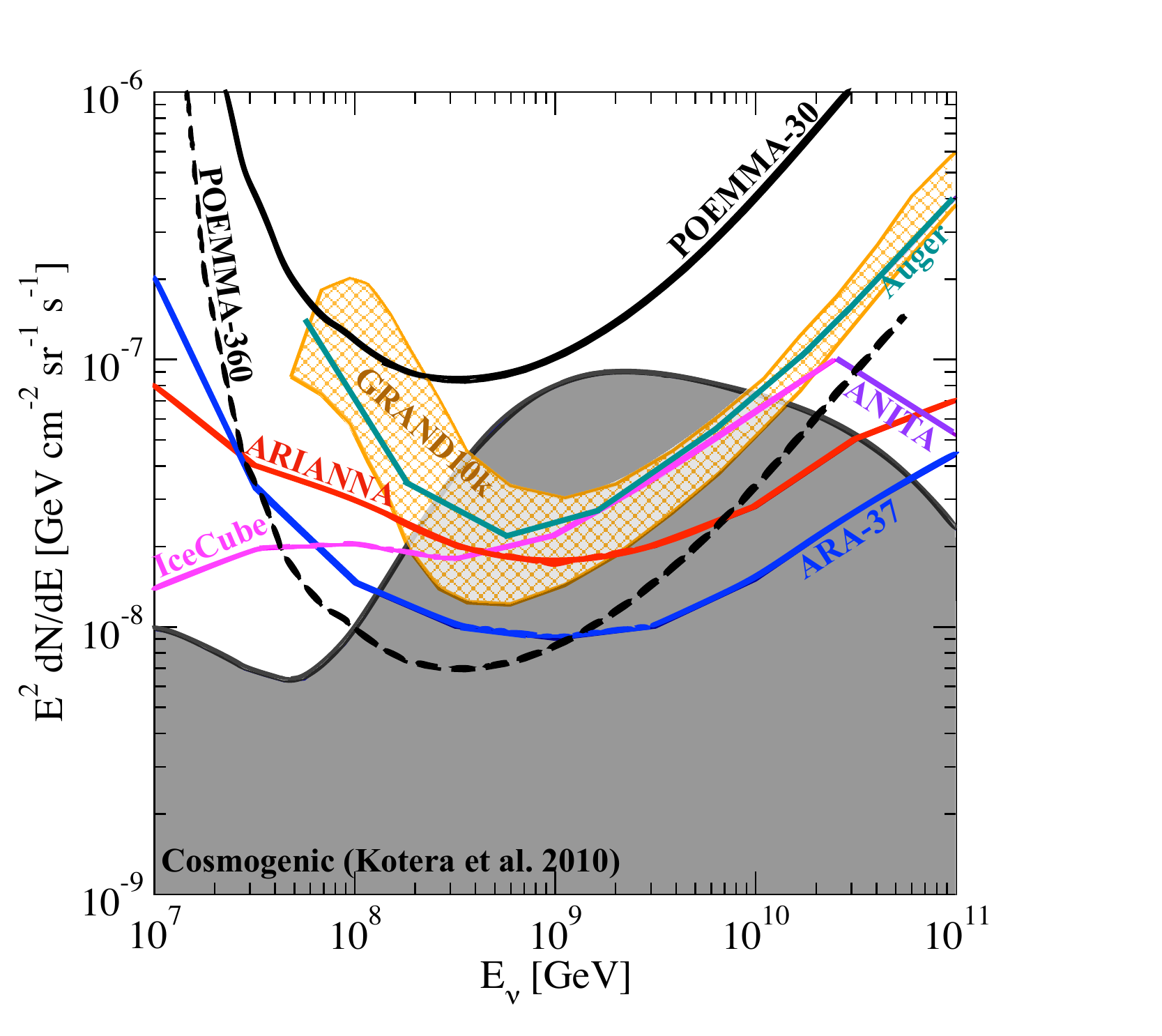} 
    \caption{{\it Left}: The  flux of tau neutrinos exiting the Earth, scaled by energy, $E_\tau$ for four elevation angles  $\beta_{\rm tr}\equiv 90^\circ-\theta_{\rm tr}$ with different 	  approximations for the cross section and energy loss. 
    {\it Right}: The sensitivity for POEMMA with $\Delta\phi = 360^\circ$ (dashed) and $30^\circ$ to the all-flavor diffuse neutrino flux scaled by energy-squared, assuming 5 years of observing with a 20\% duty cycle. A band of cosmogenic flux predictions by Kotera et al. \cite{Kotera:2010yn} is shown, along with 90\%
    CL upper limits from Auger (scaled to decade energy bins) \cite{Aab:2019auo}, IceCube \cite{Aartsen:2018vtx}, and ANITA \cite{Gorham:2019guw}, 
    and projected sensitivities for
 ARIANNA \cite{Barwick:2014pca}, ARA-37 
\cite{Allison:2011wk} and 
GRAND10k \cite{Fang:2017mhl}, scaled as necessary for the all-flavor sensitivity.}
    \label{fig:diffuse}
\end{figure}

Figure \ref{fig:diffuse} (left) shows the impact of the energy loss model and neutrino cross section for different elevation angles $\beta_{\rm tr}\equiv 90^\circ -\theta_{\rm tr}$. The quantity shown does not depend on the decay in the atmosphere or detection probability, only the tau exit probability and the predicted exit energy.

Plotted in Fig. \ref{fig:diffuse} (left) is the tau energy times the flux of upgoing taus given an input isotropic diffuse flux of neutrinos predicted by Kotera et al. \cite{Kotera:2010yn}. This model describes the flux of neutrinos from cosmic ray interactions with the cosmic background radiation, assuming uniform source evolution, a mixed cosmic ray composition and a maximum proton energy of $10^{11}$ GeV. The solid histograms show tau
flux as a function of energy, scaled by energy, for the ALLM electromagnetic energy loss. The dashed histograms show BDHM energy loss. Both solid and dashed histograms are evaluated with $\sigma_{\rm SM}$. The full band includes the minimum and maximum values allowing for the ALLM and BDHM extrapolations for $\sigma_{\nu N}$. As the elevation angle $\beta_{\rm tr}$ gets larger, the chord length through the Earth increases. Neutrino attenuation becomes more important, so there are more discrepancies between  histograms with the same energy loss but different cross section model inputs.
The larger impact at large $\beta_{\rm tr}$ doesn't translate to the effective aperture because 
most of the effective aperture comes from small $\beta_{\rm tr}$.

We use the ALLM model as our default for calculating the tau electromagnetic energy loss, with the standard model parton distribution function based $\sigma_{\rm SM}$ neutrino-nucleon cross section. The right panel of Fig. \ref{fig:diffuse} shows our predicted sensitivity for POEMMA with $\Delta\phi = 360^\circ$ (black dashed line) and $30^\circ$ (solid black line) to the all-flavor diffuse neutrino flux scaled by energy-squared. We have assumed 5 years of observation time with a $20$\% duty cycle. To be competitive in diffuse flux measurements, POEMMA's expansion to 
$\Delta\phi = 360^\circ$ is necessary, as shown in the comparison with 90\% confidence level upper limits from Auger \cite{Zas:2017xdj}, IceCube \cite{Aartsen:2018vtx},  and ANITA 
\cite{Gorham:2019guw}, and projected sensitivities of ARIANNA \cite{Barwick:2014pca}, ARA-37 
\cite{Allison:2011wk} and 
GRAND10k \cite{Fang:2017mhl}.

\begin{figure}[t]
	\begin{center}
          \includegraphics[width=0.45\columnwidth]{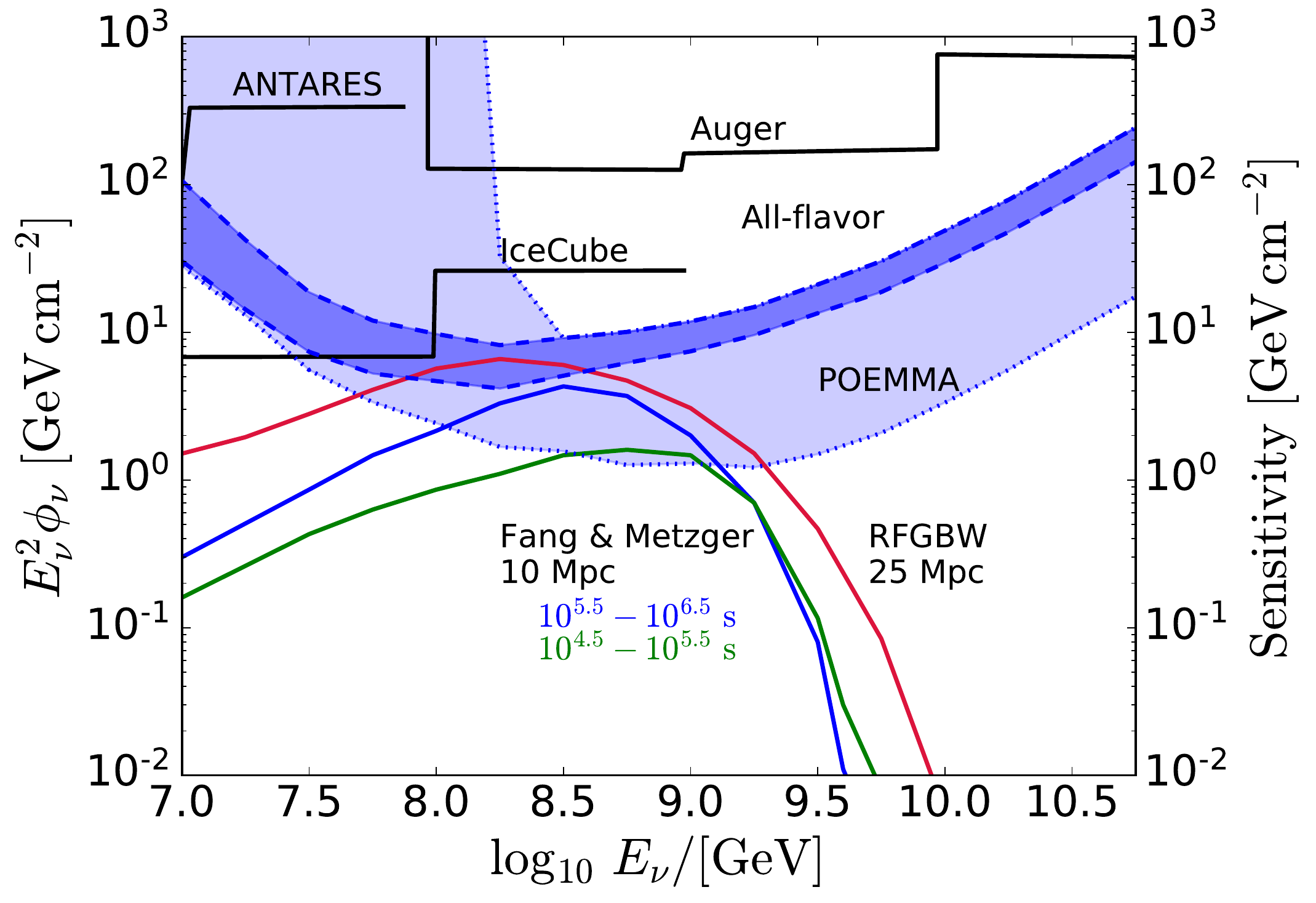}   
	\includegraphics[width=0.45\columnwidth]{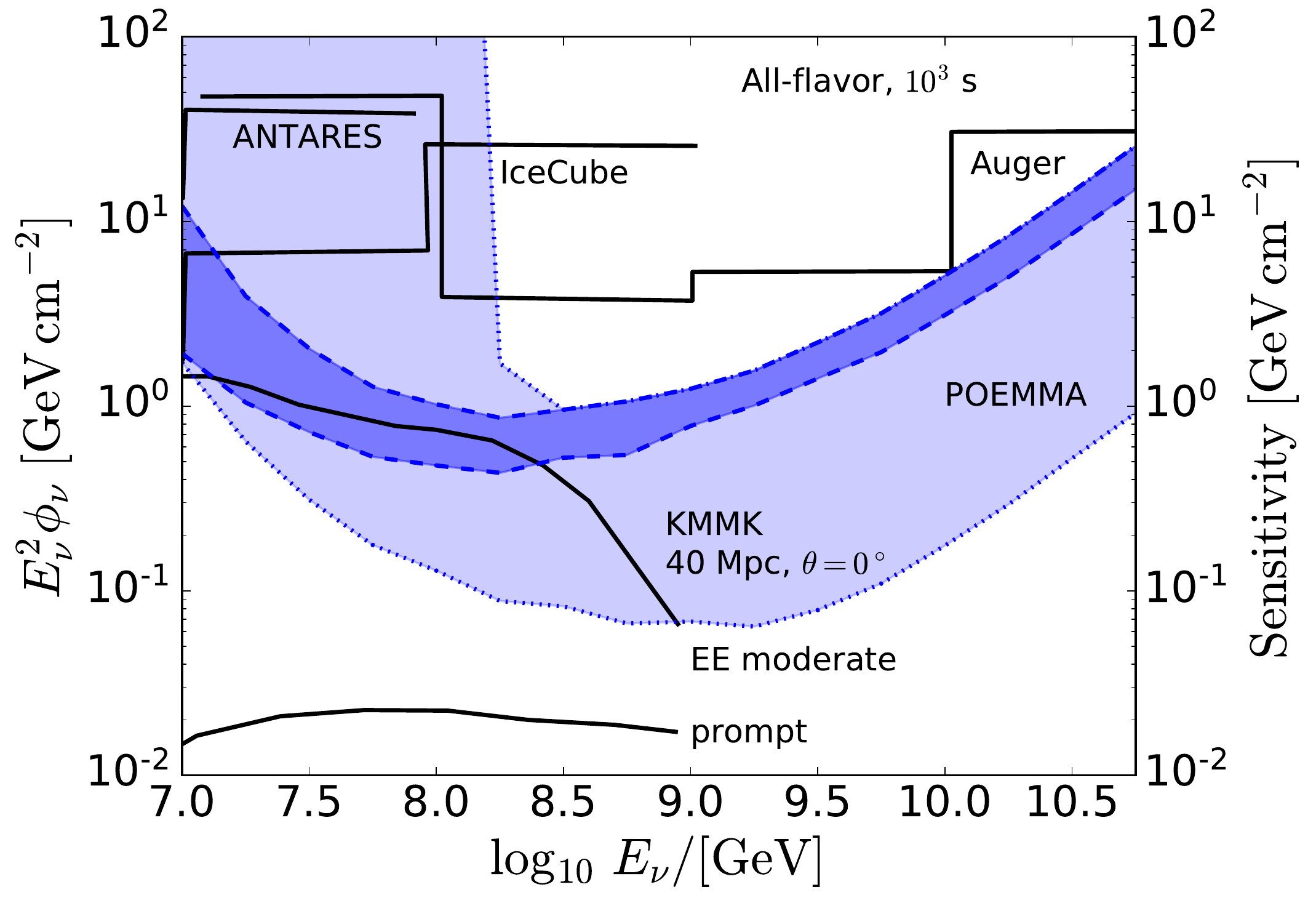}	
                 
	\end{center}
	\vspace{-0.75 cm}
	\caption{All-flavor sensitivity scaled by $E_\nu^2$, as a function of $E_\nu$. The dark blue band shows the sensitivity for a large portion of the sky at a given time, while the lighter blue region shows other viewing locations. The left figure is for long bursts, averaged over 380 days, including the effects of the Sun and Moon. The right figure shows the best all-flavor sensitivity to short bursts ($10^3$ s).  Models by Fang and Metzger \cite{Fang:2017tla}, Rodrigues et al. (RFGBW) \cite{Rodrigues:2017fmu}  and Kimura et al. (KMMK) \cite{Kimura:2017kan} are shown. See text for discussion of IceCube, Auger and ANTARES 90\% CL upper limits, scaled to 3 flavors \cite{ANTARES:2017bia}. }
	\label{fig:pointsource}	
\end{figure}

POEMMA has an advantage over other instruments for transient sources due to the fact that is a satellite-based intstrument. With an orbital period of $T_s=95$ min, an orbital plane at an angle $\xi_i=28.5^\circ$ relative to the Earth's polar axis, and a precession period of $T_p=54.3$ days, over a few months, the whole sky will be visible to POEMMA \cite{Guepin:2018yuf}. 
Figure \ref{fig:pointsource} shows POEMMA's all-flavor sensitivity to long ($\sim 10^6$ s, left) and short ($10^3$ s, right) bursts. The dark blue band shows the sensitivity for a large portion of the sky at a given time, while the lighter blue region shows special regions of better or worse sensitivity that depends on source location. The left figure, averaged over 380 days of viewing, includes a de-rating factor of order $0.3$ on average due the effects of the Sun and the Moon on observations by POEMMA \cite{Guepin:2018yuf,Venters:2019}. The right figure shows the best all-flavor sensitivity to short bursts ($10^3$ s) and does not include the effects of the Sun and the Moon. If the source is not in a favorable position during the short burst, POEMMA will not have any sensitivity, but if the Earth comes between the source and POEMMA, the instrument's fast slewing capability will result in the sensitivities shown in the right panel. Examples of long burst fluences by Fang and Metzger for a neutron star-neutron star merger scaled to 10 Mpc  \cite{Fang:2017tla} and  a blazar flare model proposed by
Rodrigues, Fedynitch, Gao, Bonioli and Winter (RFGBW) scaled to 25 Mpc \cite{Rodrigues:2017fmu} are shown.  Also plotted is an example of a short neutrino burst from a short duration gamma ray burst, the Kimura, Murase, M\'esz\'aros and Kiuchi (KMMK) \cite{Kimura:2017kan} prediction for the all-flavor fluence for extended emission and prompt emission, scaled to 40 Mpc, for on-axis viewing ($\theta = 0^\circ$).
Finally, in the left (right) plot, IceCube, Auger and ANTARES 90\% confidence level upper limits are scaled to 3 flavors from their results for 14 days after ($\pm 500$ s around) the binary neutrino star merger GW170817 \cite{ANTARES:2017bia}. POEMMA's better sensitivity for short bursts, should the source be in a viewable location, comes from the fact that the time averaged effective area is larger for short bursts: $10^3$ s is shorter than the orbital period which is relevant to the long bursts.

Given the effective area of POEMMA, one can calculate the maximum luminosity distance accessible to POEMMA for a variety of models of transient sources. Details of our evaluation appear in Ref. \cite{Venters:2019}. 
We find, for example, that the Lunardini and Winter tidal disruption event long burst models \cite{Lunardini:2016xwi} with supermassive black hole masses on the order of $10^5 M_\odot-5\times 10^6 M_\odot$ are accessible out to $\sim 100$ Mpc for 1 neutrino event in POEMMA. Long bursting blazar flares in the RFBGW model \cite{Rodrigues:2017fmu} are accessible out to $\sim 43$ Mpc. The extended emission short gamma ray burst model of KKMK \cite{Kimura:2017kan}
would produce $1$ neutrino event at POEMMA for a source distance of $\sim 117$ Mpc.

Refinements of the diffuse and transient neutrino source sensitivities are underway. Theoretical inputs to the neutrino interactions and tau energy loss give variations on the order of $\pm 30-40\%$ in the sensitivity for $E_{\nu_\tau}=10^9$ GeV where POEMMA has its best constraints.  Modeling of air showers, the impact of cloud cover and other variables are being reviewed and improved \cite{Krizmanic:2019icrc}. A NASA APRA funded software package is in development to provide a computational tool for tau neutrino induced upward air showers 
\cite{Krizmanic:2019icrc}.

\acknowledgments
We thank Roopesh Ojha, Elizabeth Hays and our  colleagues
of  the  Pierre  Auger  and  POEMMA  collaborations
for  valuable  discussions.    This  work  is  supported
in  part  by  US  Department  of  Energy  grant  DE-SC-0010113, NASA Grant 17-APRA17-0066, NASA
awards  NNX17AJ82G  and  80NSSC18K0464,  and
the  U.S.  National  Science  Foundation  (NSF  Grant
PHY-1620661).


\begin{thebibliography}{99}

\bibitem{Kotera:2011cp} 
  K.~Kotera and A.~V.~Olinto,
  Ann.\ Rev.\ Astron.\ Astrophys.\  {\bf 49}, 119 (2011)
  [arXiv:1101.4256 [astro-ph.HE]].

\bibitem{Anchordoqui:2018qom} 
  L.~A.~Anchordoqui,
  Phys.\ Rep.\  {\bf 801}, 1 (2019)
  [arXiv:1807.09645 [astro-ph.HE]].



\bibitem{Aartsen:2018vtx} 
  M.~G.~Aartsen {\it et al.} [IceCube Collaboration],
  Phys.\ Rev.\ D {\bf 98}, 
  062003 (2018)
  [arXiv:1807.01820 [astro-ph.HE]].

 
\bibitem{Sanguineti:2019pkv} 
  M.~Sanguineti [ANTARES and KM3NeT Collaborations],
  Universe {\bf 5}, 
  65 (2019).


\bibitem{Allison:2011wk} 
  P.~Allison {\it et al.},
  Astropart.\ Phys.\  {\bf 35}, 457 (2012)
  [arXiv:1105.2854 [astro-ph.IM]].

\bibitem{Barwick:2014pca} 
  S.~W.~Barwick {\it et al.} [ARIANNA Collaboration],
  Astropart.\ Phys.\  {\bf 70}, 12 (2015)
  [arXiv:1410.7352 [astro-ph.HE]].

\bibitem{Aab:2019auo} 
  A.~Aab {\it et al.} [Pierre Auger Collaboration],
  arXiv:1906.07422 [astro-ph.HE].
 
\bibitem{Zas:2017xdj} 
  E.~Zas [Pierre Auger Collaboration],
  PoS ICRC {\bf 2017}, 972 (2018).
  
\bibitem{Abreu:2012zz} 
  P.~Abreu {\it et al.} [Pierre Auger Collaboration],
  Astrophys.\ J.\  {\bf 755}, L4 (2012)
  [arXiv:1210.3143 [astro-ph.HE]].


\bibitem{Gorham:2019guw} 
  P.~W.~Gorham {\it et al.} [ANITA Collaboration],
  Phys.\ Rev.\ D {\bf 99}, no. 12, 122001 (2019)
  [arXiv:1902.04005 [astro-ph.HE]].


\bibitem{IceCube:2018dnn} 
  M.~G.~Aartsen {\it et al.} [IceCube, Fermi-LAT, MAGIC, AGILE, ASAS-SN, HAWC, H.E.S.S., INTEGRAL, Kanata, Kiso, Kapteyn, Liverpool Telescope, Subaru, Swift, NuSTAR, VERITAS and VLA/17B-403 Collaborations],
  Science {\bf 361}, 
  eaat1378 (2018)
  [arXiv:1807.08816 [astro-ph.HE]].


\bibitem{Otte:2018uxj} 
  A.~N.~Otte,
  Phys.\ Rev.\ D {\bf 99}, 
  083012 (2019)
  [arXiv:1811.09287 [astro-ph.IM]].

\bibitem{Fang:2017mhl} 
  K.~Fang {\it et al.},
  PoS ICRC {\bf 2017}, 996 (2018)
  [arXiv:1708.05128 [astro-ph.IM]].
  
\bibitem{Alvarez-Muniz:2018bhp} 
  J.~Alvarez-Mu\~niz {\it et al.} [GRAND Collaboration],
  arXiv:1810.09994 [astro-ph.HE].


\bibitem{Fargion:2000iz} 
  D.~Fargion,
  Astrophys.\ J.\  {\bf 570}, 909 (2002)
  [astro-ph/0002453].

\bibitem{Bertou:2001vm} 
  X.~Bertou, P.~Billoir, O.~Deligny, C.~Lachaud and A.~Letessier-Selvon,
  Astropart.\ Phys.\  {\bf 17}, 183 (2002)
  [astro-ph/0104452].

  
\bibitem{Feng:2001ue} 
  J.~L.~Feng, P.~Fisher, F.~Wilczek and T.~M.~Yu,
  Phys.\ Rev.\ Lett.\  {\bf 88}, 161102 (2002)
  [hep-ph/0105067].

\bibitem{Halzen:1998be} 
  F.~Halzen and D.~Saltzberg,
  Phys.\ Rev.\ Lett.\  {\bf 81}, 4305 (1998)
  [hep-ph/9804354].


\bibitem{Reno:2019jtr} 
  M.~H.~Reno, J.~F.~Krizmanic and T.~M.~Venters,
  arXiv:1902.11287 [astro-ph.HE].
 
\bibitem{Olinto:2017xbi} 
  A.~V.~Olinto {\it et al.},
  PoS ICRC {\bf 2017}, 542 (2018)
  [arXiv:1708.07599 [astro-ph.IM]].

\bibitem{Olinto:2019icrc}  
A.~V.~Olinto {\it et al.},
  PoS ICRC {\bf 2019}, 378 (2019).
  
\bibitem{Venters:2019} 
  T. M. Venters, M.~H.~Reno, J.~F.~Krizmanic, L. A. Anchordoqui, C. Gu\'epin and A.~V.~Olinto,
  arXiv:1906.07209 [astro-ph.HE].

\bibitem{Guepin:2018yuf} 
  C.~Gu\'epin, F.~Sarazin, J.~Krizmanic, J.~Loerincs, A.~Olinto and A.~Piccone,
  JCAP {\bf 1903}, 
  021 (2019)
  [arXiv:1812.07596 [astro-ph.IM]].
 
    
 
\bibitem{Motloch:2013kva} 
  P.~Motloch, N.~Hollon and P.~Privitera,
  Astropart.\ Phys.\  {\bf 54}, 40 (2014)
  [arXiv:1309.0561 [astro-ph.IM]].
 
\bibitem{Alvarez-Muniz:2017mpk} 
  J.~Alvarez-Mu\~niz, W.~R.~Carvalho, A.~L.~Cummings, K.~Payet, A.~Romero-Wolf, H.~Schoorlemmer and E.~Zas,
  Phys.\ Rev.\ D {\bf 97}, 
  023021 (2018)
  Erratum: [Phys.\ Rev.\ D {\bf 99}, 
  069902 (2019)]
  [arXiv:1707.00334 [astro-ph.HE], arXiv:1901.08498 [astro-ph.HE]].
  
\bibitem{Jeong:2017mzv} 
  Y.~S.~Jeong, M.~V.~Luu, M.~H.~Reno and I.~Sarcevic,
  Phys.\ Rev.\ D {\bf 96}, 
  043003 (2017)
  [arXiv:1704.00050 [hep-ph]].
  
\bibitem{Abramowicz:1991xz} 
  H.~Abramowicz, E.~M.~Levin, A.~Levy and U.~Maor,
  Phys.\ Lett.\ B {\bf 269}, 465 (1991).
\bibitem{Abramowicz:1997ms} 
  H.~Abramowicz and A.~Levy,
  [hep-ph/9712415].

\bibitem{Block:2014kza} 
  M.~M.~Block, L.~Durand and P.~Ha,
  Phys.\ Rev.\ D {\bf 89}, 
  094027 (2014)
  [arXiv:1404.4530 [hep-ph]].
  
 
\bibitem{PalomaresRuiz:2005xw} 
  S.~Palomares-Ruiz, A.~Irimia and T.~J.~Weiler,
  Phys.\ Rev.\ D {\bf 73}, 083003 (2006)
  [astro-ph/0512231].


\bibitem{Kotera:2010yn} 
  K.~Kotera, D.~Allard and A.~V.~Olinto,
  JCAP {\bf 1010}, 013 (2010)
  [arXiv:1009.1382 [astro-ph.HE]].


\bibitem{Fang:2017tla} 
  K.~Fang and B.~D.~Metzger,
  Astrophys.\ J.\  {\bf 849}, 153 (2017)
  [arXiv:1707.04263 [astro-ph.HE]].
  
\bibitem{Rodrigues:2017fmu} 
  X.~Rodrigues, A.~Fedynitch, S.~Gao, D.~Boncioli and W.~Winter,
  Astrophys.\ J.\  {\bf 854}, 
  54 (2018)
  d
  [arXiv:1711.02091 [astro-ph.HE]].

\bibitem{Kimura:2017kan} 
  S.~S.~Kimura, K.~Murase, P.~M\'esz\'aros and K.~Kiuchi,
  Astrophys.\ J.\  {\bf 848}, 
  L4 (2017)
  [arXiv:1708.07075 [astro-ph.HE]].

\bibitem{ANTARES:2017bia} 
  A.~Albert {\it et al.} [ANTARES and IceCube and Pierre Auger and LIGO Scientific and Virgo Collaborations],
  Astrophys.\ J.\  {\bf 850}, 
  L35 (2017)
  [arXiv:1710.05839 [astro-ph.HE]].
  
\bibitem{Lunardini:2016xwi} 
  C.~Lunardini and W.~Winter,
  Phys.\ Rev.\ D {\bf 95}, 
  123001 (2017)
  [arXiv:1612.03160 [astro-ph.HE]].

  
 \bibitem{Krizmanic:2019icrc}
 J. F. Krizmanic {\it et al.},  PoS ICRC {\bf 2019}, 936 (2019).
 

\end{thebibliography}
\end{document}